\documentclass[conference]{IEEEtran}

\usepackage{amsmath,amssymb,amsfonts}
\usepackage{graphicx}
\usepackage{textcomp}
\usepackage{xcolor}
\usepackage{fancyhdr}
\usepackage{natbib}
\usepackage{breakurl}
\usepackage[breaklinks]{hyperref}
\usepackage[linesnumbered,ruled,vlined]{algorithm2e}
\usepackage{lastpage}
\def\BibTeX{{\rm B\kern-.05em{\sc i\kern-.025em b}\kern-.08em
    T\kern-.1667em\lower.7ex\hbox{E}\kern-.125emX}}

\SetCommentSty{mycommfont}

\SetKwInput{KwInput}{Input}                
\SetKwInput{KwOutput}{Output}              

\begin{document}

\title{\emph{Positional}-Unigram Byte Models for Generalized TLS Fingerprinting}

\author{\IEEEauthorblockN{\textbf{Hector A. Valdez}}
\IEEEauthorblockA{Intel Labs\\
hector.a.valdez@intel.com}
\and
\IEEEauthorblockN{\textbf{Sean McPherson}}
\IEEEauthorblockA{Intel Labs \\
sean.mcpherson@intel.com}
}

\maketitle 

\begin{abstract}
We use \emph{positional}-unigram byte models along with maximum likelihood for generalized TLS fingerprinting and empirically show that it is robust to cipher stunting. Our approach creates a set of \emph{positional}-unigram byte models from client hello messages. Each \emph{positional}-unigram byte model is a statistical model of TLS client hello traffic created by a client application or process. To fingerprint a TLS connection, we use its client hello, and compute the likelihood as a function of a statistical model. The statistical model that maximizes the likelihood function is the predicted client application for the given client hello. Our data driven approach does not use side-channel information and can be updated on-the-fly. We experimentally validate our method on an internal dataset and show that it is robust to cipher stunting by tracking an \emph{unbiased} $f_{1}$ score as we synthetically increase randomization.

\end{abstract}

\begin{IEEEkeywords}
TLS fingerprinting, maximum likelihood, unigram, cipher stunting
\end{IEEEkeywords}

\section{\textbf{Introduction}}
The Transport Layer Security (TLS) protocol, formerly Secure Sockets Layer (SSL), establishes an encrypted connection between two hosts on a network. For increased security, majority of web applications now use TLS when establishing internet connections \citep{googlereport}. Similarly, a rise in the use of TLS for malicious applications has been observed \citep{sophos}. These malicious applications use TLS to hide behind encryption to propagate malware, communicate with infected devices via command and control servers, and ex-filtrate data, among others.   

Before network traffic between two hosts is encrypted, there are clear text messages sent. Threat hunters rely on tools, such as JA3 \citep{Althouse2019}, to create fingerprints of these clear text messages and build databases that map fingerprint to application, process, or library name. This helps detect and identify many, but not all, malicious applications attempting to establish a secure connection. However, the manner in which these fingerprints are created can be exploited. Malicious actors have developed counter measures to avoid being detected and exploit the vulnerabilities of well-known fingerprint methods. 

One counter measure that has been deployed by attackers to evade detection is cipher stunting \citep{AlamaiBlog2019}. Attackers use cipher stunting to randomize a client hello cipher suite list, which induces a fingerprint that might not be in the defenders database. If these unknown fingerprints cannot be mapped to a known application, it becomes difficult to know if they are benign or malicious without further investigation of the connection attributes. Manually investigating and sorting through all of these unknown connections becomes unmanageable and time intensive. Therefore, it is imperative to develop automated tools that alleviate the burden on the threat hunter, addresses current limitations of common industry tools, and provides improved network defense.

In \citep{Anderson2020}, the authors use a knowledge base and destination context to generalize TLS fingerprinting. They form a knowledge base with TLS fingerprint strings along with destination IP addresses, ports, and server names, to disambiguate between processes using a Naive Bayes classifier. However, we seek a solution that does not rely on header information since it can be spoofed or modified, causing these features to appear random or similar to benign processes. 

We introduce a classification method that is a step towards generalized TLS fingerprinting and robust to changes in a client hello message. Our method uses \emph{positional}-unigram byte models built from client hello messages, where each \emph{positional}-unigram byte model is assigned a label dependent on a pre-defined labeling process. Consequently, each \emph{positional}-unigram byte model captures statistical information from client hello messages that are related by some attribute(s), conditioned by the labeling process. During inference, we compute the likelihood of a client hello message belonging to each \emph{positional}-unigram byte model and predict client hello membership by selecting the \emph{positional}-unigram byte model that maximizes a likelihood function. An additional benefit of our procedure is that updating these \emph{positional}-unigram byte models with new information is simple and can be performed on-the-fly.  

As mentioned in \citep{Anderson2020}, carefully selecting fields in a client hello message to construct a fingerprint string can result in many client applications mapping to the same TLS fingerprint. Unlike JA3, we incorporate all bytes of a client hello message, except fields that contain irrelevant information, e.g. random bytes. Nevertheless, we use JA3 to assist in automatically labeling our dataset since we did not have ground truth labels for the client hello samples. Therefore, we regard our automatic labeling process as producing \emph{quasi}-labels, which approximate the true label assignment of our dataset. Inference, on the other hand, is completely reliant on \emph{positional}-unigram byte models and the procedure we describe in \ref{ourmethod}.

Our contributions include: 
\begin{itemize}
\item a generalized TLS fingerprinting method which incorporates statistical information from known client applications to classify client hello messages.
\item a study of our method verses the commonly used JA3 method applied to cipher stunted client hello messages.
\item a study of our method using all bytes in a client hello versus using JA3 bytes
\end{itemize}

\section{\textbf{Background}}
For completeness, we present fundamental material related to TLS handshake, client hello message, and packet data.

\subsection{\textbf{TLS Handshake}}
\begin{figure}[htbp]
\centerline{\includegraphics[scale=0.4]{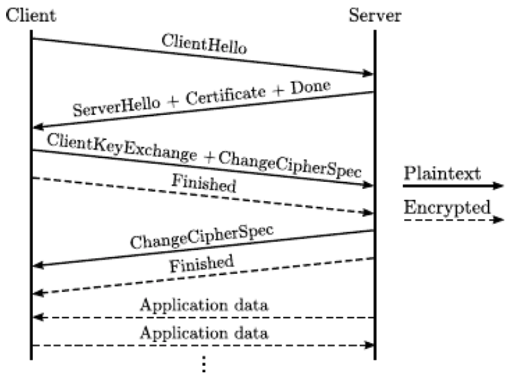}}
\caption{TLS handshake between Client and Server. From \citep{Husak2015}.}
\label{fig1}
\end{figure}

The TLS handshake is a negotiation between Client and Server via a series of messages. The Client is the entity that initiates the negotiation. The initial communication is a three-way handshake where the Client sends a connection request SYN to a Server. If the Server is listening and available, the Server sends SYN/ACK to acknowledge the connection request SYN. Once the Client receives the SYN/ACK, the Client sends a final ACK as the acknowledgment for the SYN/ACK. This three-way communication kicks off the main TLS handshake. Fig.~\ref{fig1} shows a notional representation of the main TLS handshake which includes the initial client hello from the Client, followed by the response from the Server called server hello, and ending with the key exchange before encryption begins.

\subsection{\textbf{Client Hello}}
The constituent fields of a client hello are: \emph{protocol version}, \emph{random bytes}, \emph{session ID}, \emph{list of cipher suites}, \emph{list of compression methods}, and \emph{list of extensions}. The fixed fields present in a client hello message are the \emph{protocol version}, \emph{random bytes}, 
\emph{session ID}, \emph{list of cipher suites}, and \emph{list of compression methods}. The \emph{list of extensions} are variable fields which change depending on the client application generating the client hello.

\begin{figure}[htbp]
\centerline{\includegraphics[scale=0.5]{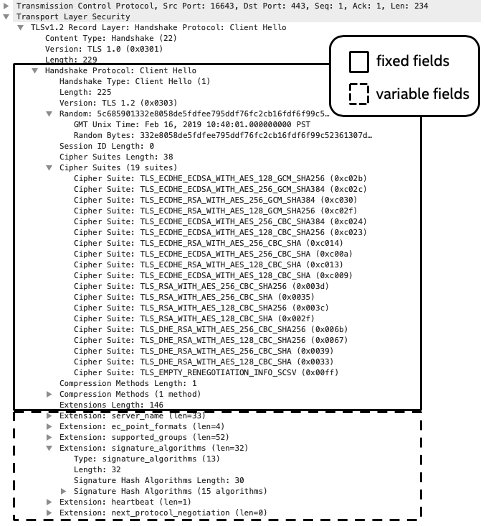}}
\caption{client hello message viewed in Wireshark. Fixed fields are enclosed in solid rectangle and variable fields are enclosed in dotted rectangle. This particular client hello, sent a list of 19 cipher suites to the Server. The Server will subsequently cross 
reference this list to its own list and select the most secure cipher suite.}
\label{fig2}
\end{figure}

Certain fields in a client hello contain random bytes including, \emph{random bytes}, \emph{session ID}, and infrequently an early \emph{key share} when TLS 1.3 is invoked. We show an example client hello viewed in Wireshark in Fig.~\ref{fig2}, which illustrates the ordering of fields and shows a unique list of cipher suites from the \emph{cipher suite} field. 

The client hello and server hello are the messages in a handshake that establish the security level of communication and agree on what cipher suite and compression method to use for encryption. However, our method does not use any bytes from the server hello. Only the client hello is used and contents of subsequent messages in a TLS handshake are ignored. Segments of the client hello worth understanding are the following:

\subsubsection{\textbf{List of Cipher Suites}}
Each cipher suite contains a set of algorithms that are used for encryption. They can be grouped into three distinct categories: key exchange, bulk encryption, and message authentication code. For a detailed list of cipher suites refer to Appendix A Section 5 in \citep{Dierks2008}.

\subsubsection{\textbf{List of Compression Methods}}
The algorithms that compress and decompress packet payloads sent with encryption. For the detailed standards of TLS compression refer to \citep{Hollenbeck2004}.

\subsubsection{\textbf{List of Extensions}}
Extensions add functionality to TLS without modifying the protocol itself. For a detailed list of extensions and a description of each extension's purpose refer to \citep{Eastlake2011}.

\subsection{\textbf{Packet Data}}
\begin{figure}[htbp]
\centerline{\includegraphics[scale=0.5]{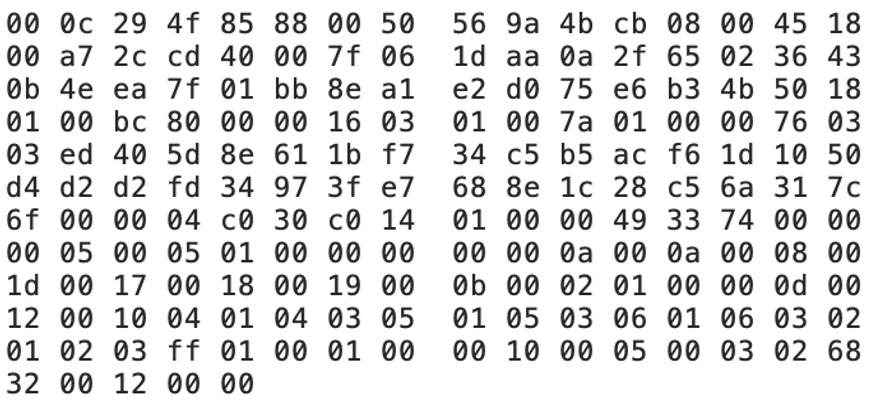}}
\caption{client hello message in hexadecimal form.}
\label{fig3}
\end{figure}

Raw client hello data is in binary, but gets converted to human-friendly hexadecimal, i.e. base-16. Fig.~\ref{fig3} displays an example of a client hello in hexadecimal form.

\subsubsection{\textbf{Hexadecimal}}
Hexadecimal has sixteen distinct symbols. These symbols begin with the traditional base-10 digits, 0-9, followed by the first six letters of the English alphabet, a-f. Each of these symbols is a human-friendly digit in base-16 and has a 4-bit binary representation called a nibble, or half a byte. A pair of hexadecimal digits form one byte. Therefore, the number of unique bytes, or possible hexadecimal pairings, is 256.

\subsubsection{\textbf{Hexadecimal to Decimal}}
\begin{figure}[htbp]
\centerline{\includegraphics[scale=.45]{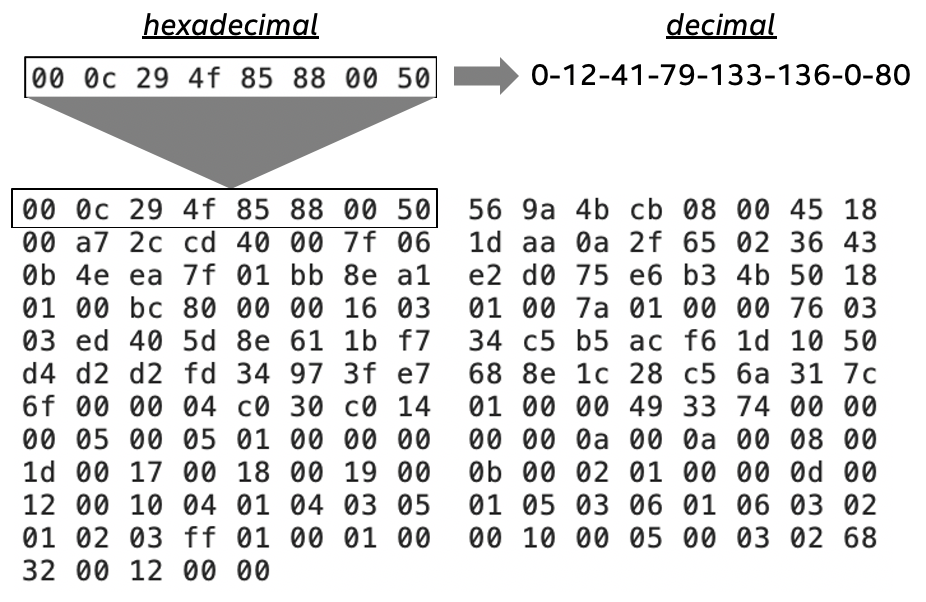}}
\caption{Top line of client hello packet converted from hexadecimal to decimal.}
\label{fig4}
\end{figure}

Since there are exactly 256 unique hexadecimal pairings, each can be converted to a decimal representation, with each hexadecimal pairing being assigned a value in the range 0-255. Fig.~\ref{fig4} demonstrates the first line of a client hello packet being converted from hexadecimal to decimal.

\section{\textbf{The JA3 Method}}
JA3 is a TLS fingerprinting method that operates specifically on client hello messages. JA3 extracts specific client hello fields and converts them to decimal values. The information extracted for the JA3 method are: \emph{protocol version}, \emph{list of cipher suites}, \emph{extension codes}, \emph{elliptic curve}, and \emph{elliptic curve point format}. The decimal values are concatenated into a string that is ordered according to the list previously mentioned. Each field value is delimited by ``," and each value in a field is delimited by ``-". This string is then digested by the MD5 hashing function and produces a unique 16-byte fingerprint, i.e. 32 hexadecimal valued fingerprint. If there are any empty fields in a client hello, then those corresponding fields in the decimal string are also left empty. The interested reader can read more about the JA3 method in \citep{Althouse2019}.

There is a follow-on method called JA3S that is a fingerprinting tool for server hello messages and is used jointly with JA3 to reduce false positives. However, JA3 is the only technique we are interested in for the purposes of this work and evaluate our model against this tool. Before we present our procedure and describe how we evaluate it, we discuss limitations of JA3 and how they motivated the design of our method.

\subsection{\textbf{discarded information}}
Specific JA3 fields were meticulously chosen by its creators to produce fingerprints for commonly used client applications and known malicious software. This leaves some client hello data out of processing altogether, e.g. compression method and signature algorithm, so there is a possibility that discarded data might help in some situations and contain useful information needed during inference. Therefore, we include all client hello data for our procedure, except client hello header, any randomly generated fields, e.g. random bytes, and server name. 

\subsection{\textbf{collisions}}
We make the distinction between a hash function collision and a JA3 hash collision. The former can happen due to the nature of the hash function, but the possibility of this is extremely low and we assume this will not occur in practise. The later is more likely since the JA3 hash is dependent on the JA3 decimal string, which might be the same for two different client applications \citep{Bakker2018}. For this reason, we incorporate more information in our procedure to reduce false positives and false negatives.

\subsection{\textbf{lack of robustness}}
Since a specific JA3 decimal string induces a unique JA3 hash, changing one decimal value in the string will yield a different JA3 hash. For example, a simple cipher suite swap that fundamentally does not change the function of a client hello, will induce a different JA3 hash and may not be in the Network Defender's JA3 hash database. We know from \citep{AlamaiBlog2019} that this method of cipher stunting is commonly used by attackers. Having a lack of robustness and being sensitive to the manipulation of bytes in a client hello message is detrimental to a network defense system. As a consequence, this was the primary motivation of developing a generalized TLS fingerprinting method that is robust to perturbations in the client hello message.

\section{\textbf{\emph{Positional}-Unigram Byte Models from Client Hello Messages}}\label{ourmethod}
Consider the probability of observing a unique sequence of bytes $x_{1},...,x_{m}$, where $x_{i}$ is a byte value at position $i$ in the sequence. The joint probability of $x_{1},...,x_{m}$ is the product of conditional probabilities

\begin{equation}\label{jointP}
\begin{split}
p(x_{1},...,x_{m}) &= p(x_{1}) p(x_{2} | x_{1}) \dotsm p(x_{m} | x_{1},..., x_{m-1}) \\
                   &= \prod_{k=1}^{m}{p(x_{k} | x_{1},...,x_{k-1})}. 
\end{split}
\end{equation}

For $m$ large, calculating the conditional probability $p(x_{m} | x_{1},..., x_{m-1})$ is infeasible, therefore, we rely on approximations of these conditional probabilities to estimate the joint probability. We make a Markov assumption and assume the Markov property holds for sequence windows of length $n$. Then the joint probability of $x_{1},...,x_{m}$
is approximated as follows

\begin{equation}\label{approx_joint}
\begin{split}
p(x_{1},...,x_{m}) &\approx p(x_{1}) p(x_{2} | x_{1}) \dotsm p(x_{m} | x_{m-(n-1)},...,x_{m-1}) \\
                   &= \prod_{k=1}^{m}{p(x_{k} | x_{k-(n-1)},...,x_{k-1})}.
\end{split}
\end{equation}
The joint probability defined in Eq.~\ref{approx_joint} is an \emph{n}-gram model. In the extreme case, we let $n=1$, which is the unigram model

\begin{equation}
\begin{split}
p(x_{1},...,x_{m}) &\approx p(x_{1}) p(x_{2}) \dotsm p(x_{m}) = \prod_{k=1}^{m}{p(x_{k})} 
\end{split}
\end{equation}
where each $p(x_{k})$ is an unconditional probability. Estimating unconditional probabilities from data is accomplished by dividing the frequency count of byte occurrence by the total sum of all possible byte occurrences. We modify this unigram byte model to account for position of byte occurrence in a sequence, while preserving the ease of computing unconditional probabilities.

Now, consider a \emph{positional}-unigram byte model composed of positional dependent unconditional probabilities, such that at every position $i$ we have 

\begin{equation}
\begin{split}
1=\sum_{b=0}^{255}{p_{i}(x=b)},
\end{split}
\end{equation}
where byte $b$ is in decimal form.

Each probability $p_{i}(x=b)$ is calculated by dividing the frequency count of byte $b$ occurring at position $i$ by the total sum of frequency counts at position $i$. Then the \emph{positional}-unigram byte model is 

\begin{equation}
\begin{split}
p(x_{1},...,x_{m}) &\approx p_{1}(x_{1}) p_{2}(x_{2}) \dotsm p_{m}(x_{m}) = \prod_{k=1}^{m}{p_{k}(x_{k})}.
\end{split}
\end{equation} 

Our method is presented in two parts. The first part describes the construction of \emph{positional}-unigram byte models that we use for inference, and the second part describes the inference procedure that determines client hello class membership, i.e., predicting the client application that generated the client hello.

\subsection{\textbf{creating positional-unigram byte models}}
The procedure to create \emph{positional}-unigram byte models will now be described.

\subsubsection{\textbf{label}}
Given a training dataset $T$ of client hello messages, we label each sample $\mathbf{x}$ using JA3. We begin by extracting JA3 specific fields from $\mathbf{x}$ and convert these fields into their decimal equivalents. Then these decimal values are concatenated with dashes in between values. The next step is to transform this string of concatenated decimal values into an MD5 hash. Then the MD5 hash gets mapped to a label using a JA3 repository composed of key-value pairs. The JA3 repository is a hash table with hashes as keys and application names as values. If there is a hash not in the repository, then it gets mapped to $Unknown$. Let $L$ be the set of all labels in our repository and let $l \in L$ be a label. It is important to have many \emph{(hash, l)} pairs in the repository so that there is a high portion of client hello messages labeled, otherwise most will get mapped to \emph{Unknown} and will be rendered useless for our training stage. Our JA3 labels were derived from multiple online JA3 hash repositories. The full database consists of key-value pairs \emph{(MD5 hash, label)} from: \citep{sslblacklist}, \citep{ja3er}, \citep{tna}, \citep{salesforce}, \citep{brotherston}. Some example JA3 labels are:
\emph{'Adware'}, \emph{'Chrome Version 60/61.0.3163'}, \emph{'Chrome/71.0.3578.80 Linux 64-bit'}, \emph{'Firefox 63.0'}. More than one MD5 hash may map to the sample label.

\subsubsection{\textbf{remove irrelevant bytes}}\label{random_bytes}
For every labeled client hello message, we remove \emph{random bytes}, \emph{session ID}, and  \emph{key share} since they all contain random information that does not assist in prediction. We also remove \emph{server name} since this can be easily altered and we do not want our models to rely on this field during inference. Once all irrelevant bytes have been removed, we filter any client hello duplicates, i.e. keep only one copy of a unique byte sequence. We filter out duplicates since they add computational overhead in \ref{step3} and get normalized out in \ref{step4}.

\subsubsection{\textbf{count byte occurrences}}\label{step3}
For every label $l \in L$, create a set of frequency counts $\{f_{i}^{(l)}\}_{i=1,...,m_{l}}$, where $m_{l}=\text{max} \left( \{ \text{length}(x) | x \in J_{l}\} \right)$ and $J_{l}$ is the set of client hello messages assigned label $l$. Each frequency count vector at position $i$ is defined as $f_{i}^{(l)} = \left( o_{i,0}, o_{i,1}, ..., o_{i,255} \right)^{(l)}$ and is a running count of byte occurrence for each position $i$ over each byte value $b$, represented by $o_{i,b}$, where $b=0,...,255$.
If we stack the set of frequency counts $\{f_{i}^{(l)}\}_{i=1,...,m_{l}}$ ordered by position, we form a frequency count matrix $\mathbf{F}^{(l)}$ of size $m_{l} \times 256$. 
For reasons that involve how we perform inference, we apply additive smoothing to these frequency counts. So we initialize each frequency count at a value $0<\delta<1$. For our experiments we set $\delta = 1 \times 10^{-8}$.

\begin{figure}[htbp]
\centerline{\includegraphics[scale=0.27]{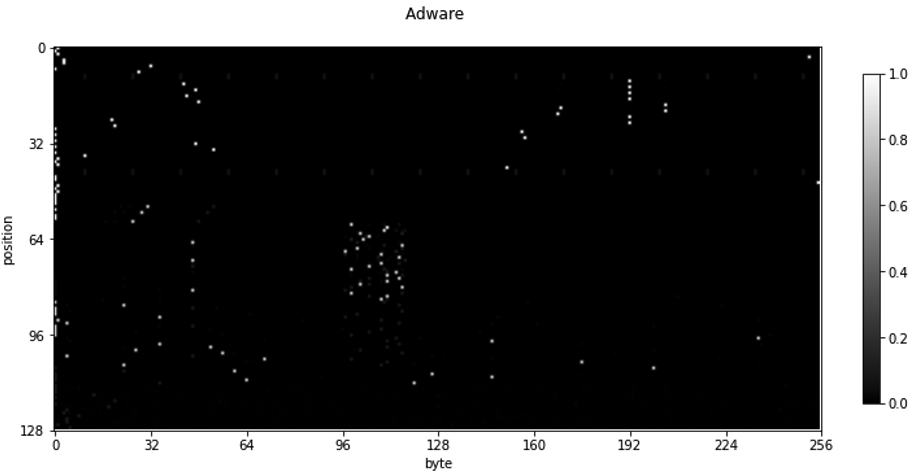}}
\caption{Visualization of a $J_{Adware}$ \emph{positional}-unigram byte model built from 3,024 \emph{Adware} client hello messages. Only the first 128 of 448 positions are displayed.}
\label{fig7}
\end{figure}

\subsubsection{\textbf{normalize frequency counts}}\label{step4}
To create a \emph{positional}-unigram byte model, we divide each frequency count vector $f_{i}^{(l)}$ at position $i$ by the total sum $s_{i}^{(l)}=\sum_{b=0}^{255}{o_{i,b}}$ at position $i$, for label $l \in L$. So,

\begin{equation}
\begin{split}
P_{i}^{(l)} &= \frac{f_{i}^{(l)}}{s_{i}^{(l)}} \\
            &= \frac{\left( o_{i,0}^{(l)}, o_{i,1}^{(l)}, ..., o_{i,255}^{(l)} \right)}{s_{i}^{(l)}} \\
            &= \left( \frac{o_{i,0}^{(l)}}{s_{i}^{(l)}}, \frac{o_{i,1}^{(l)}}{s_{i}^{(l)}}, ..., \frac{o_{i,255}^{(l)}}{s_{i}^{(l)}} \right) \\
            &= \left( p_{i,0}^{(l)}, p_{i,1}^{(l)}, ..., p_{i,255}^{(l)} \right)
\end{split}
\end{equation}
is the probability vector of a \emph{positional}-unigram byte model at position $i$, for label $l \in L$. Therefore, each $p_{i,b}$ is the probability of byte $b$ occurring at position $i$.

The equivalent matrix transformation of $\mathbf{F}^{(l)}$ is computed by calculating the sum of each row in the frequency count matrix $\mathbf{F}^{(l)}$, and then, dividing each row by its corresponding row sum. The transformed matrix $\mathbf{P}^{(l)}$ will be called the \emph{positional}-unigram byte matrix of size $m_{l} \times 256$.

Now, $\mathbf{P}^{(l)}$ contains statistical information from all client hello messages in $J_{l}$. Each row $i$ of the $J_{l}$ \emph{positional}-unigram byte matrix $\mathbf{P}^{(l)}$, can be interpreted as a categorical distribution
over all possible byte values $b$. Fig.~\ref{fig7} shows the statistical information that emerges from a $J_{\emph{Adware}}$ \emph{positional}-unigram byte matrix.

\begin{algorithm}
\DontPrintSemicolon
    
    \KwInput{client hello messages}
    \KwOutput{\emph{positional}-unigram byte models}
    \KwData{dataset $T$, client hello $\mathbf{x} \in T$, JA3 labels $L$}
    
    \tcp{partition client hello messages into distinct $J_{l}$ sets}
    \ForAll{$\mathbf{x} \in T$}
    {
        label = JA3($\mathbf{x}$) \\
        \If{\textnormal{label} $\in L$}
        {
            place $\mathbf{x}$ in $J_{\text{label}}$
        }
        \Else{discard $\mathbf{x}$}
    }

    \tcp{create \emph{positional}-unigram byte models}
    \ForAll{$l \in L$}
    {
        \ForAll{$\mathbf{x} \in J_{l}$}
        {
            $\mathbf{x} =$  remove\_irrelevant\_information($\mathbf{x}$) \\
        }
        $J_{l} =$ remove\_duplicate\_messages($J_{l}$)

        $\{f_{i}^{(l)}\}_{i=1,...,m_{l}} \longleftarrow J_{l}$ \\
        $\mathbf{F}^{(l)} \longleftarrow \{f_{i}^{(l)}\}_{i=1,...,m_{l}}$

        $\mathbf{P}^{(l)}$ = normalize($\mathbf{F}^{(l)}$)
    }
    \Return{$\{ \mathbf{P}^{(l)} \}_{\forall l \in L}$}
\caption{create \emph{positional}-unigram byte models}
\end{algorithm}

An additional benefit of our method is that we can easily update any frequency count on-the-fly, which entails performing the JA3 method on a client hello and then determining if it can be added to one of the $J_{l}$ sets. Then we update the corresponding frequency count matrix $\mathbf{F}^{(l)}$, followed by updating $\mathbf{P}^{(l)}$. If JA3 induces a new label for the new client hello, then we simply create a new $\mathbf{F}^{(l)}$ and $\mathbf{P}^{(l)}$.

\subsection{\textbf{inference}}
Each \emph{positional}-unigram byte model is a statistical model formed from training data that were generated from an application or process. Now, consider a hypothesis set containing all statistical models $\{ \mathbf{P}^{(l)} \}_{\forall l \in L}$. Let $\mathbf{x}$ be a client hello with irrelevant bytes extracted as described in section \ref{random_bytes}. We define the likelihood $\mathcal{L}$ of an observed client hello $\mathbf{x}$ as the function of a statistical model

\begin{equation}
\mathcal{L}(\mathbf{P}^{(l)} | \mathbf{x}) = \prod_{k=0}^{m} p_{k}(x_{k})
\end{equation} 
and find the statistical model that maximizes the likelihood

\begin{equation}
\arg\max_{\forall l \in L} \mathcal{L}(\mathbf{P}^{(l)} | \mathbf{x}).
\end{equation}

For practical reasons, we make some algebraic adjustments since multiplying probabilities is not ideal and the statistical models are of variable length. Therefore, the likelihood function we actually use during inference is the mean log-likelihood

\begin{equation}
\begin{split}
\hat{l} &= \arg\max_{\forall l \in L} \frac{1}{K} \text{log} \mathcal{L}(\mathbf{P}^{(l)} | \mathbf{x}) \\
      &= \arg\max_{\forall l \in L} \frac{1}{K} \sum_{i=1}^{K}\text{log}(p_{i,x_{i}}^{(l)}) 
\end{split}
\end{equation}

where 

\begin{equation}
\begin{split}
x_{i} &= \text{byte value at position } i \text{ in } \mathbf{x} \\
K &= \text{min}(h, m_{l}), \\
h &= \text{length of client hello } \mathbf{x}, \\
m_{l} &= \text{number of rows in } \mathbf{P}^{(l)}. \\
\end{split}
\end{equation}

Label $\hat{l}$ corresponds to the statistical model that maximizes the likelihood and is the predicted label for client hello $\mathbf{x}$.

\begin{algorithm}
\DontPrintSemicolon
    
    \KwInput{client hello and \emph{positional}-unigram byte models }
    \KwOutput{label of predicted client application}
    \tcp{preprocess client hello}
    Extract irrelevant bytes from client hello to form $\mathbf{x}$ \\

    \tcp{find statistical model that maximizes the mean log-likelihood}
    $h = \text{length}(\mathbf{x})$ \\
    $score = - \infty$ \\
    label = $\emptyset$ \\
    \ForAll{$l \in L$}
    {
        $m_{l} = $ number of rows in $\mathbf{P}^{(l)}$\\
        $K = \text{min}(h, m_{l})$ \\
        \If{score $< \frac{1}{K}$ \text{log} $\mathcal{L}(\mathbf{P}^{(l)} | \mathbf{x})$}
        {
            $score = \frac{1}{K} \text{log} \mathcal{L}(\mathbf{P}^{(l)} | \mathbf{x})$ \\
            label $= l$
        }
    }
    \Return{\textnormal{label}}

\caption{inference}
\end{algorithm}

\subsubsection{\textbf{inference time and space complexity}}
It is obvious to see that the inference procedure on a client hello has constant space complexity $\mathcal{O}(1)$ and linear time complexity $\mathcal{O}(M)$, where $M=\text{max}(\{\text{min}(h, m_{l})\}_{l \in L})$.

\section{\textbf{Experimentation}}
We now present two experiments we performed using an internal dataset.

\subsection{\textbf{Internal TLS dataset}}
The TLS dataset used in our experiments was collected internally and is derived from one network. The entire packet capture (pcap) was approximately 25 gigabytes (GB) of which 11 GB was TLS traffic. The TLS traffic was split into eight pcap files \emph{(split-1.pcap, split-2.pcap, split-3.pcap, split-4.pcap, split-5.pcap, split-6.pcap, split-7.pcap, split-8.pcap)} ranging from 1.2$-$1.6 GB in size. Each file included both client hello and server hello messages; however, only client hello data was used for our experiments. In Table \ref{datasettotals} we list total number of client hello samples contained in each pcap (\textbf{total samples}), total number of client hello samples with \emph{Unknown} client hello messages removed (\textbf{without} \emph{Unknown}), and total number of client hello samples with \emph{Unknown} and duplicate samples removed (\textbf{unique}). To arrive at total number of \textbf{unique} samples, we take total number of samples \textbf{without} \emph{Unknown} and extract all irrelevant bytes from each client hello message. Then we filter out all duplicate samples. Once these duplicate samples are removed, we are left with \textbf{unique} samples, which are the samples we use in our experiments.

We consider each client application in the dataset as a class and discovered that there is significant class imbalance. There are a total of $48,906$ unique samples across $121$ classes. The top 4 classes with the most samples account for approximately 75\% of the dataset and the top 14 classes with the most samples account for approximately 95\% of the dataset. The class most represented had 21,129 samples, in contrast to the 12 least represented classes with only 2 samples each.

\begin{table}[htbp]
\caption{}
\begin{center}
\begin{tabular}{|c|c|c|c|}
\hline
& & & \\
\textbf{split} & \textbf{total samples} & \textbf{without }\emph{Unknown} & \textbf{unique$^{\mathrm{*}}$} \\
\hline
\emph{split-1} & 127,818 & 123,694 & 7,451 \\
\emph{split-2} & 109,877 & 106,260 & 6,319 \\
\emph{split-3} & 96,411 & 93,268 & 5,772 \\
\emph{split-4} & 106,636 & 103,372 & 6,156 \\
\emph{split-5} & 103,027 & 100,101 & 5,957 \\
\emph{split-6} & 104,571 & 101,444 & 6,494 \\
\emph{split-7} & 89,535 & 87,089 & 5,409 \\
\emph{split-8} & 92,044 & 89,495 & 5,348 \\
\hline
\multicolumn{4}{l}{$^{\mathrm{*}}$\emph{Unknown} and duplicate samples filtered out}
\end{tabular}
\label{datasettotals}
\end{center}
\end{table}

\subsection{\textbf{performance metric}}
For the multi-class classification task, we report an \emph{unbiased} $f_{1}$ score, which weights all classes equally to account for class imbalance. When computing the $f_{1}$ score of a class, we consider positive samples as members of the class and negative samples as members of all other classes. We use the definition of $f_{1}$ to compute the score for each class

\begin{equation}
\begin{split}
f_{1} &= 2 \cdot \frac{\text{precision} \cdot \text{recall}}{\text{precision} + \text{recall}}
\end{split}
\end{equation}
where
\begin{equation}
\begin{split}
\text{precision} &= \frac{\text{true positives}}{\text{true positives} + \text{false positives}} 
\end{split}
\end{equation}
and
\begin{equation}
\begin{split}
\text{recall} &= \frac{\text{true positives}}{\text{true positives} + \text{false negatives}}. \\
\end{split}
\end{equation}

Now, let $c$ be a class in the set of all classes $\mathbf{C}$. Then we define
\begin{equation}
\begin{split}
\emph{unbiased}\text{ }f_{1} &= \frac{1}{|\mathbf{C}|} \sum_{c \in \mathbf{C}} f_{1}^{(c)}
\end{split}
\end{equation}
where $|\mathbf{C}|$ is the number of classes.

\subsection{\textbf{Experiment 1a - Ordered Cipher Suite Permutations}}\label{exp1a}
For each client hello under test, we synthetically modify its \emph{cipher suite list} by randomly selecting a cipher suite in the list and swap the selected cipher suite with the next cipher suite in the list. We call this an \emph{ordered cipher suite permutation}. We use client hello messages containing cipher suite lists of length 2 or greater for obvious reason.

\subsubsection{\textbf{Data}}\label{exp1a_data}
We perform $k$-fold cross validation on the \emph{internal TLS dataset} using each pcap as a split, so $k=8$. For example, 1 fold would use \emph{split-1}, \emph{split-2}, \emph{split-3}, \emph{split-4}, \emph{split-5}, \emph{split-6}, and \emph{split-7} as training data and use \emph{split-8} as validation data. For the \emph{ordered cipher suite permutation} experiment we introduce one random swap, so we perform 30 trials for each fold to place confidence bounds around the mean. In total, we performed 240 trials across all folds.

\subsubsection{\textbf{Results}}\label{exp1a_results}
Since we compare our method to JA3, we report \emph{ordered cipher suite permutations} with and without GREASE values \citep{Benjamin2018}. As mentioned in \citep{Althouse2019}, JA3 ignores GREASE values to ensure that client applications using these values are identified with a single JA3 hash. A subset of cipher suites happen to be GREASE values, so when we permute a GREASE value with an adjacent non-GREASE value, JA3 will yield the same hash for the permuted and non-permuted cipher suite list. We report results of \emph{ordered cipher suite permutations} with GREASE values in Table \ref{cipher_suite_tab1} and \emph{ordered cipher suite permutations} without GREASE values in Table \ref{cipher_suite_tab2}. We report the unbiased $f_{1}$ score with a $95 \%$ confidence interval.

For Table \ref{cipher_suite_tab1} and \ref{cipher_suite_tab2}, notice that if there are no permutations, JA3 has perfect performance since JA3 is used as the automated labeling tool for our dataset. However, once a client hello undergoes an \emph{ordered cipher suite permutation}, JA3 experiences catastrophic failure since unknown hashes are produced after a random cipher suite swap. In contrast, our method of maximum likelihood experiences minimal performance degradation if there is a random cipher suite swap present. 

\subsection{\textbf{Experiment 1b - Ordered Cipher Suite Permutations using JA3 bytes}}\label{exp1b}
We conduct the same experiment as described in \ref{exp1a}, but instead of using all the bytes in a client hello message we use the JA3 decimal string equivalents. Meaning, we build frequency counts in the same fashion as described in \ref{step3} using the decimal values of \emph{protocol version}, \emph{list of cipher suites}, \emph{extension codes}, \emph{elliptic curve}, and \emph{elliptic curve point format}.

\subsubsection{\textbf{Data}}
Exactly the same as in \ref{exp1a_data}.

\subsubsection{\textbf{Results}} 
We see similar trends as described in \ref{exp1a_results}, except using JA3 decimal string equivalents performs slightly better than using all bytes from a client hello message. This result is most likely due to the fact that frequency counts built with with JA3 decimal string equivalents are closer to the \emph{quasi}-labels produced from the automated labeling tool.

\begin{table}[htbp]
\caption{}
\begin{center}
\begin{tabular}{|c|c|c|c|}
\hline
&\multicolumn{3}{|c|}{\textbf{\textit{Ordered Cipher Suite Permutation}} w/ GREASE} \\
\cline{2-4} 
\textbf{method} & \textbf{\textit{permutation$^{\mathrm{*}}$}} & \textbf{all bytes} & \textbf{JA3 bytes only}\\
\hline
\emph{JA3} & none & $1.0$ & $1.0$\\
 & yes & $.0180 \pm .0007$ & $.0188 \pm .0009$ \\
\hline
\emph{maximum} & none & $.9285 \pm .009$ & $.9314 \pm .0064$ \\
\emph{likelihood} & yes & $.9206 \pm .0016$ & $.9330 \pm .0013$ \\
\hline
\multicolumn{4}{l}{$^{\mathrm{*}}$only one ordered cipher suite permutation}
\end{tabular}
\label{cipher_suite_tab1}
\end{center}
\end{table}

\begin{table}[htbp]
\caption{}
\begin{center}
\begin{tabular}{|c|c|c|c|}
\hline
&\multicolumn{3}{|c|}{\textbf{\textit{Ordered Cipher Suite Permutation}} w/o GREASE} \\
\cline{2-4} 
\textbf{method} & \textbf{\textit{permutation$^{\mathrm{*}}$}} & \textbf{all bytes} & \textbf{JA3 bytes only} \\
\hline
\emph{JA3} & none & $1.0$ & $1.0$\\
 & yes & $.0$ & $.0$ \\
\hline
\emph{maximum} & none & $.9223 \pm .0067$ & $.9314 \pm .0064$ \\
\emph{likelihood} & yes & $.9205 \pm .0016$ & $.9332 \pm .0013$ \\
\hline
\multicolumn{4}{l}{$^{\mathrm{*}}$only one ordered cipher suite permutation}
\end{tabular}
\label{cipher_suite_tab2}
\end{center}
\end{table}

\subsection{\textbf{Experiment 2a - Random Cipher Suite Permutations}}\label{exp2a}
For each client hello under test, we synthetically modify its \emph{cipher suite list} by randomly selecting a fraction of cipher suites in the list and permute the selected cipher suites. We call this a \emph{random cipher suite permutation}. We do a grid search in the interval [0.1, 1.0] with steps of 0.1 as fraction values for random permutations. For example, a value of 0.1 means 10 percent of the \emph{cipher suite list} is permuted. We use client hello messages containing cipher suite lists of length 2 or greater for obvious reason.

\subsubsection{\textbf{Data}}\label{exp2a_data}
We perform $k$-fold cross validation on the \emph{internal TLS dataset} using each pcap as a split, similar to \ref{exp1a_data}. For the \emph{random cipher suite permutation} experiment we introduce random permutations on a fraction of the \emph{cipher suite list}, so we perform 4 trials per fraction value to place confidence bounds around the mean at a specific fraction value. Consequently, we performed 32 trials for each fraction value.

\subsubsection{\textbf{Results}}
\begin{figure}[htbp]
\centerline{\includegraphics[scale=0.55]{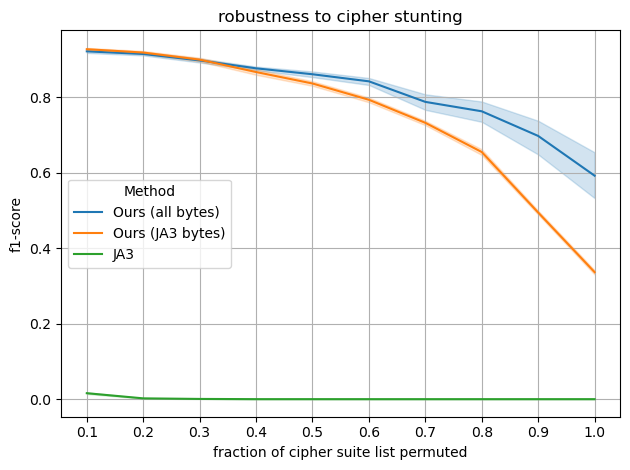}}
\caption{The maximum likelihood of \emph{positional}-unigram byte models, given a client hello, is robust to cipher stunting.}
\label{f1_avg}
\end{figure}

In this experiment, we do not remove client hello messages that had GREASE values permuted, since it is more akin to a realistic environment. Figure \ref{f1_avg} shows maximum likelihood using \emph{positional}-unigram byte models built with \textbf{all bytes} is robust to a high fraction of random cipher suite permutations, compared to the catastrophic failure of JA3. This indicates that our method maintains robustness across \emph{positional}-unigram byte models of various levels of sample support.

\begin{figure}[htbp]
\centerline{\includegraphics[scale=0.55]{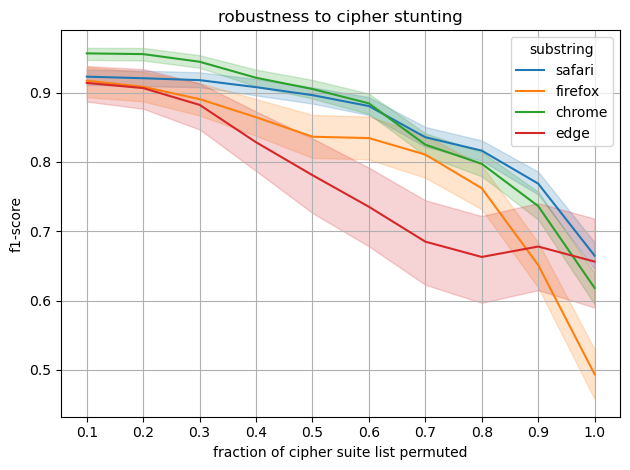}}
\caption{We perform a substring search across all class labels and average the $f_{1}$ performance metric of any classes where the substring is found in the class label. We show the performance of substring searches: "safari", "chrome", "firefox", "edge", which are browser names. Performance results are of maximum likelihood using \emph{positional}-unigram byte models built with \textbf{all bytes}.}
\label{f1_avg_browsers}
\end{figure}

\begin{figure}[htbp]
\centerline{\includegraphics[scale=0.55]{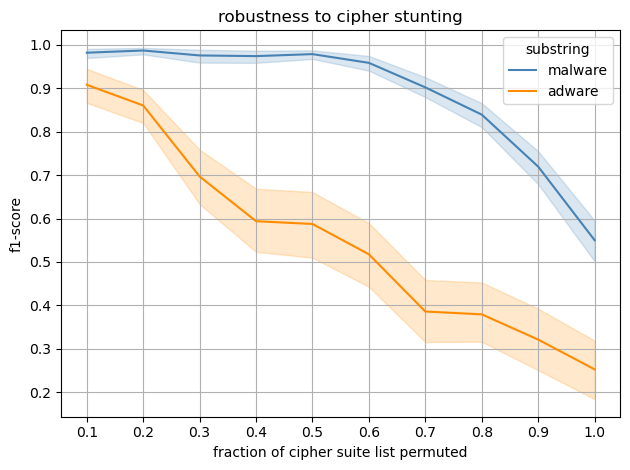}}
\caption{We perform a substring search across all class labels and average the $f_{1}$ performance metric of any classes where the substring is found in the class label. We show the performance of substring searches: "malware", "adware", which are considered malicious names. Performance results are of maximum likelihood using \emph{positional}-unigram byte models built with \textbf{all bytes}.}
\label{f1_avg_process}
\end{figure}

To understand how maximum likelihood performs on a subset of classes, or super-set, conditioned on a specific substring, we perform a substring search across all class labels and average the $f_{1}$ performance metric of any classes where the substring is found in the class label. In Figure \ref{f1_avg_browsers}, we show browser keyword searches. In Figure \ref{f1_avg_process}, we show process type keyword searches.

\subsection{\textbf{Experiment 2b - Random Cipher Suite Permutations using JA3 bytes}}
We conduct the same experiment as described in \ref{exp2a}, but instead of using all bytes in a client hello message we use the JA3 decimal string equivalents similar to \ref{exp1b}. 

\subsubsection{\textbf{Data}}
Exactly the same as in \ref{exp2a_data}.

\subsubsection{\textbf{Results}}
In this experiment, we do not remove client hello messages that had GREASE values permuted, similar to \ref{exp2a}. Figure \ref{f1_avg} shows maximum likelihood using \emph{positional}-unigram byte models built with \textbf{JA3 bytes} is not as robust to a high fraction of random cipher suite permutations, compared to its \textbf{all bytes} counterpart. We also see that the variance in the \emph{unbiased} $f_{1}$ score is smaller compared to \textbf{all bytes}. Since \emph{positional}-unigram byte models were built with \textbf{JA3 bytes} in this experiment, the \emph{cipher suite list} tends to make up more of the information content. Therefore, when a high fraction of the \emph{cipher suite list} is randomly permuted, performance degrades more rapidly compared to \emph{positional}-unigram byte models built with \textbf{all bytes}.

\section{\textbf{Conclusion}}
We have shown that maximum likelihood using \emph{positional}-unigram byte models is robust to cipher stunting. Our results also confirm that JA3 catastrophically fails in the presence of cipher stunting. Lastly, we also show that building \emph{positional}-unigram byte models with all bytes in a client hello performs better than building the same models with JA3 bytes when comparing them against an \emph{unbiased} $f_{1}$ score. 

In practice, network defenders use an assortment of tools to detect and mitigate malicious traffic on a network, and we have presented another practical tool that can be deployed to assist the network defender. Our method can be used as a supplemental tool when JA3 encounters an \emph{Unknown} hash.

\section{\textbf{Future Work}}
As discussed, we used JA3 as an automated labeling tool to label our dataset which might have errors. Ideally, we would have ground truth labels for each process or client application we create a \emph{positional}-unigram byte model. Therefore, curating a dataset with ground truth labels would yield more faithful performance results than the approximations presented.

In this paper, we explored using unigram models as our statistical models, however, we could extend this to bigram and trigram models which would incorporate conditional probabilities. 

As mentioned earlier, we did not include server hello data to disambiguate between processes or applications, although, there might be useful information present in the server hello that can help prediction.

{
    \small
    \bibliographystyle{unsrtnat}
    \bibliography{pubmodel}
}


\end{document}